# Echoes of a Squeezed Oscillator

R. Merlin and A. Bianchini*

*The Harrison M. Randall Laboratory of Physics, University of Michigan, Ann Arbor, Michigan 48109-1040, USA*

Parametric coupling is commonly used to prepare oscillators in a time-varying squeezed state in which the variance of the coordinate or momentum may dip below that of the equilibrium thermal or quantum ground state during a fraction of the period. We show here that pulses applied to drive parametrically an inhomogeneously broadened set of harmonic oscillators result in a spontaneous recovery of coherence, manifesting itself as a novel type of echoes, similar to those exhibited by an ensemble of spins aligned by a magnetic field, when excited by properly designed electromagnetic pulses. Such echoes, of classical or quantum nature, are expected to arise in the squeezing of linear systems of various sorts and, in particular light and vibrational modes. Unlike existing methods for the generation of spin echoes, squeezed echoes do not require the use of resonant excitation.

In quantum mechanics, squeezing is used to describe non-stationary states for which the uncertainty of a particular operator can become smaller than that for the vacuum [1]. Although the term was originally introduced in quantum electrodynamics [2,3] and, more generally, to describe bosonic systems such as lattice modes [4,5], magnons [6], and localized vibrations [7,8,9], it has also been applied to spin states [10,11] as well as classical oscillators [12,13,14]. The reduction of noise below the thermal- or shot-noise limit is the main motivation for numerous proposals and experimental realizations of squeezing [1,2,4,6,7,10,12,13,14,15,16] including the use of squeezed light in the detection of gravitational waves [17].

Unrelated to squeezing, spin echoes refer to the refocusing of the magnetization that results from the application of electromagnetic pulses to a heterogeneous ensemble of ordered spins [18,19], an effect that is central to magnetic resonance imaging [20]. Closely related to spin echoes are photon echoes, which involve the interaction of laser pulses with few-level systems [21] and are widely used in quantum optics and quantum information science [22].

Spin and photon echoes involve the application of *resonant* pulses [23]. In this letter, we show that echoes can arise in a bosonic system under *off-resonance* conditions, as the impulsive parametric excitation of a dephased, heterogeneous set of squeezed oscillators can lead to a spontaneous resurgence of coherence, similar to that exhibited by a collection of few-level systems. We note that echoes caused by phase resetting have been recently observed in oscillatory chemical reactions [24].

Squeezed states can be produced using different methods, which all exploit either a particular nonlinear property of the system one wishes to squeeze or, more commonly, of its interaction with an ancilla [4,7,10,16,25,26]. Here, we focus on squeezing resulting from frequency control, for which the relevant Hamiltonian is that of a frequency-driven parametric oscillator

$$H = P^2/2 + \left[\Omega^2 - 2g(t)\right]Q^2/2 \qquad (1)$$

where $Q$ is the displacement from equilibrium, $P$ is the canonical momentum and $\Omega$ is the frequency. We are interested in the case where the frequency is modified by a sequence of two pulses applied at $t=0$ and $t=\Delta$. Provided the duration of the pulses is small compared with the period of oscillations, we approximate

$$g = \mu_1\delta(t) + \mu_2\delta(t-\Delta) \quad ; \qquad (2)$$

$\mu_1$ and $\mu_2$ are constants. We note that, other than squeezed light [15], electromagnetic pulses have been used to generate (quantum and classical) squeezed states involving various condensed-matter excitations [4,5,6,27,28,29], in molecular-like systems [7,8] and in macroscopic mechanical oscillators [16]. In most of these cases, the relevant effective Hamiltonian is of the form given by Eq. (1). In the following, we discuss separately the effect of the two-pulse sequence on the dynamics of $Q$ and $Q^2$. Note that, because the harmonic potential is quadratic in $Q$, the classical variables obey the same dynamical equations of motion as the corresponding quantum expectation values $\langle\Psi|Q|\Psi\rangle$ and $\langle\Psi|Q^2|\Psi\rangle$ ($\Psi$ is the wavefunction satisfying $H\Psi = i\hbar\partial\Psi/\partial t$).

We discuss first $Q$-echoes by concentrating on the classical problem. Let $Q = Q_0$ and $\dot{Q} = P_0$ be the position and momentum of a single oscillator just before the first pulse is applied. The pulses do not change the displacement, but introduce a sudden change in the momentum. For $0 < t < \Delta$, $Q(t) = U(t)$ where

$$U(t) = Q_0\cos\Omega t + \frac{(P_0 + 2\mu_1 Q_0)}{\Omega}\sin\Omega t \qquad (3)$$

and, for $t > \Delta$,

$$Q(t) = U(t) + \frac{2\mu_2 U(\Delta)}{\Omega}\sin\Omega(t-\Delta) \ . \qquad (4)$$

Consider now the behavior of $\bar{Q}$, where the bar denotes the mean over a particular inhomogeneous ensemble. Results for a set of $N_0$ oscillators with Lorentzian-distributed frequencies are shown in Fig. 1 (their number per unit of frequency is given by $dN/d\omega = \frac{\gamma N_0 / \pi}{(\Omega - \Omega_0) + \gamma^2/4}$). The oscillators are assumed to have all the same displacement and momentum at $t = 0$ (quantum mechanically, such a situation corresponds to oscillators prepared in a coherent state before the arrival of the first pulse). Following the initial excitation pulse, which adds the same amount to the momentum of all the oscillators, $\bar{Q}$ decays with time as different oscillators move at different periods. Mirroring the spin-echo problem [18], the second pulse, at $t = \Delta$, partially removes the inhomogeneous dephasing, and the evolution rephases coherently to produce the $\bar{Q}$-echo at time $2\Delta$.

The results in Fig. 1 can be explained simply by rewriting the second term of Eq. (4) as the sum of two terms of the form $\sin(\Omega t + \alpha)$ and (the echo signal) $\sin[\Omega(t - 2\Delta) + \alpha']$, where the phases depend on the frequency. As one integrates over all frequencies, dephasing leads, respectively, to the observed decaying oscillations with maxima at $t = 0$ and $t = 2\Delta$. Moreover, the absence of a term of the form $\sin[\Omega(t - \Delta + \alpha'')]$ explains the fact that the second pulse has no effect on $\bar{Q}$ at $t > \Delta$.

The (parametric) force acting on the oscillators is $F = 2Q[\mu_1 \delta(t) + \mu_2 \delta(t - \Delta)]$. To highlight the central role played by the frequency nonlinearity, and for the purpose of comparison, we show in the inset results for $F = \Pi_1 \delta(t) + \Pi_2 \delta(t - \Delta)$, that is, for a (linear) interaction term of the form $-QF(t)$. Unlike the parametric case, the second pulse leads to larger amplitude oscillations at $t \approx \Delta$ and there is no echo in the response.

The ensemble snapshots shown in Fig. 2 help understand the rephasing property of the echo phenomenon. After the response to the first pulse dies out, the second pulse, at $t = \Delta$, instantly transforms an ensemble that is (roughly) uniformly distributed in the circumference $P^2 + \Omega_0^2 Q^2 = P_0^2 + \Omega_0^2 Q_0^2$ into one that is distributed in an ellipse whose axes are rotated with respect to the $\Omega_0 Q - P$ axes. This ellipse rotates and dephases, before echoing at $t = 2\Delta$ when it becomes again a circle, but with a displaced origin.

For a thermal distribution of identical oscillators, $\overline{Q}_T \equiv 0$. A simple calculation using Eqs. (3) and (4), together with $\overline{P_T^2} = \Omega^2 \, \overline{Q_T^2} = k_B T$, gives the thermal variance for $t < \Delta$

$$\overline{Q_T^2} = \frac{k_B T}{\Omega^2}\left[\left(1 + 2\mu_1^2/\Omega^2\right) + \frac{2\mu_1}{\Omega}\sin 2\Omega t - \frac{2\mu_1^2}{\Omega^2}\cos 2\Omega t\right] \tag{5}$$

and the echo signal at $t \approx 2\Delta$

$$\overline{Q_T^2} = \frac{2\mu_1 \mu_2^2}{\Omega^5} k_B T \left[\sin 2\Omega(t - 2\Delta) + \frac{\mu_1}{\Omega}\cos 2\Omega(t - 2\Delta)\right] . \tag{6}$$

Here, $k_B$ is Boltzmann's constant and $T$ is the temperature. According to Eq. (5), the duty cycle for classical squeezing, that is, the fraction of the period for which $\overline{Q_T^2} < k_B T / \Omega^2$ is largest for $\mu_1 \ll \Omega$ and goes to zero when $\mu_1 \to \infty$. Because of the classical-quantum correspondence, this also applies to the squeezed vacuum state if one replaces $k_B T$ with $\hbar\Omega/2$.

Concerning the dynamics of $Q^2$, we are particularly interested in quantum states that have no classical counterpart such as the squeezed vacuum [2]. Defining $\Xi(t) = \langle \Psi | Q^2 | \Psi \rangle$, we get $\dot{\Xi} = -i\hbar + 2\langle \Psi | QP | \Psi \rangle$, $\ddot{\Xi} = 2\langle \Psi | P^2 | \Psi \rangle - 2\left[\Omega^2 - 2g(t)\right]\Xi$ and, finally,

$$\dddot{\Xi} + 4\dot{\Xi}\left[\Omega^2 - 2g(t)\right] - 4\dot{g}(t)\Xi = 0 \quad . \tag{7}$$

Free of forces, $\Xi$ oscillates with frequency $2\Omega$. Impulsive excitation leads to discontinuities in $\dot{\Xi}$ and $\ddot{\Xi}$, but not in $\Xi$. If $\Xi = \zeta$, $\dot{\Xi} = \dot{\zeta}$ and $\ddot{\Xi} = \ddot{\zeta}$ just before one of the pulses is applied, immediately after the pulse we have $\Xi = \zeta$, $\dot{\Xi} = \dot{\zeta} + 4\mu\zeta$ and $\ddot{\Xi} = \ddot{\zeta} + 4\mu\dot{\zeta} + 8\mu^2\zeta$. Assume that the oscillator is in its ground state, $\Psi_0$, for $t < 0$. Then, for $0 < t < \Delta$, $\Xi(t) = \Sigma(t)$ where

$$\Sigma(t)/\Xi_0 = \left(1 + \frac{2\mu_1^2}{\Omega^2}\right) + \frac{2\mu_1}{\Omega}\sin 2\Omega t - \frac{2\mu_1^2}{\Omega^2}\cos 2\Omega t \quad ; \tag{8}$$

$\Xi_0 = \langle \Psi_0 | Q^2 | \Psi_0 \rangle = \hbar/2\Omega$ is the variance of the zero-point motion. It is worth to point out that, after the pulse, the oscillator finds itself instantly in the squeezed state $e^{-i\mu Q^2/\hbar}\Psi_0$. Also, note that $\langle \Psi | Q | \Psi \rangle = 0$ at all times, so that $\Xi$ is always equal to the variance.

For $t > \Delta$, we get

$$\Xi(t) = \Sigma(t) + \left(\frac{\mu_2\dot{\Sigma}(\Delta) + 2\mu_2^2\Sigma(\Delta)}{\Omega^2}\right)$$
$$+ \frac{2\mu_2\Sigma(\Delta)}{\Omega}\sin 2\Omega(t-\Delta) - \left(\frac{\mu_2\dot{\Sigma}(\Delta) + 2\mu_2^2\Sigma(\Delta)}{\Omega^2}\right)\cos 2\Omega(t-\Delta) \tag{9}$$

Given that $\Sigma(\Delta)$ is of the form $A\sin(2\Omega\Delta + \varphi) + B$, $\Xi(t)$ will exhibit additional oscillations from $t = \Delta$ due to the second pulse and, in addition, an echo at $t \approx 2\Delta$. Explicitly, the echo signal is

$$\Xi_{\text{ECHO}}(t) = \frac{2\mu_1\mu_2^2}{\Omega^3}\Xi_0\left[\sin 2\Omega(t-2\Delta) + \frac{\mu_1}{\Omega}\cos 2\Omega(t-2\Delta)\right] \quad . \tag{10}$$

Other than for the constant factors, Eq. (8) and the above expression are identical to the corresponding thermal expressions, Eq. (5) and Eq. (6) (thus, as anticipated, the previous comment about the duty cycle applies to the quantum case). Again, this is a manifestation of the fact that classical oscillators also obey Eq. (7) (as does $\langle\Psi|Q|\Psi\rangle^2$, even though $\langle\Psi|Q^2|\Psi\rangle \neq \langle\Psi|Q|\Psi\rangle^2$). Reflecting the equivalence between classical and quantum probability densities [30], we observe that Eqs. (8-9) can be obtained if one takes $\overline{P_0^2} = \Omega^2 \overline{Q_0^2}$ and $\overline{P_0 Q_0} = 0$ after squaring Eqs. (3-4).

The results in Fig. 3 are for a set with a Lorentzian distribution of frequencies; each oscillator is initially in its quantum ground state. The data shows the decaying oscillations in the averaged variance $\overline{\Xi(t)}$ induced by the two pulses and, as anticipated, rephasing leading to the coherent echo at $t = 2\Delta$. The plot in the inset is for classical oscillators, all of which have $Q = Q_0$ and $P = P_0$ just before the first pulse (as mentioned earlier, the corresponding quantum state is a coherent state with $\langle\Psi|Q|\Psi\rangle = Q_0$ and $\langle\Psi|P|\Psi\rangle = P_0$). The echo here manifests itself as a dip in the variance $\overline{Q^2} - (\overline{Q})^2$. The overall behavior of $\overline{Q}$ is identical to that in Fig. 1. A simple analysis using Eqs. (3) and (4) indicates that $\overline{Q^2}$ (or the averaged $\langle\Psi|Q|\Psi\rangle^2$) exhibits three sets of $2\Omega$-oscillations starting at $t = 0$ and $t = \Delta$, and peaking at $t = 2\Delta$, which originate, respectively, from the terms $\sin^2(\Omega t + \alpha)$, $\sin(\Omega t + \alpha)\sin[\Omega(t - 2\Delta) + \alpha']$ and $\sin^2[\Omega(t - 2\Delta) + \alpha']$. The tail end of the oscillations induced by the second pulse can be seen in the inset of Fig. 3, just before the echo. Note that the evolution of the classical ensemble after the first pulse mimics that of a coherent-squeezed quantum state [31,32].

In summary, we have shown that an inhomogeneous set of parametrically driven harmonic oscillators, when impulsively excited, behave in a manner that is very similar to that of aligned spins or few-level systems under resonant pulsed electromagnetic excitation. As for the latter, the occurrence of squeezed echoes holds promise for the development of techniques to distinguish homogeneous from inhomogeneous broadening in the decay of molecular vibrations, phonons, magnons and other bosons, as well as imaging methods similar to those that rely on magnetic resonance.

# FIGURE CAPTIONS

Figure 1 – Classical oscillators. Time-dependence of $\overline{Q(t)}$ for a Lorentzian distribution of frequencies, with $\Omega_0=7.5$ and $\gamma = 0.5$. The parametric force acting on the (undamped) oscillators is $-2g(t)Q$, with $g = \mu_1\delta(t)+\mu_2\delta(t-\Delta)$ ($\Delta = 10$, $\mu_1=5$ and $\mu_2=2.5$), and the common initial conditions are $Q_0 = 25$ and $P_0 = 10$. The echo is at $t \approx 2\Delta$. Inset: Results for the two-pulse force $F = \Pi_1\delta(t)+\Pi_2\delta(t-\Delta)$ with $\Pi_1 =1$, $\Pi_2 = -100$ and $\Delta = 10$. Parameters are $Q_0 = 25$, $P_0 = 1$, $\Omega_0 = 10.0$ and $\gamma = 0.5$.

Figure 2 – Snapshots of an ensemble of $10^4$ oscillators with a Lorentzian distribution of frequencies. The pulses leading to the echo at $t = 2\Delta$ strike at $t = 0$ and $t = \Delta$. Conditions and parameters are the same as in the main Figure 1.

Figure 3 – Time dependence of the mean variance $\overline{\Xi(t)}$ for a set of quantum oscillators with Lorentzian distributed frequencies; see Eqs. (8) and (9). The parameters of the distribution are $\Omega_0 = 7.5$ and $\gamma = 0.5$. The oscillators are initially in the ground state and are acted upon by the two-pulse force $-2g(t)Q$ with $g = \mu_1\delta(t)+\mu_2\delta(t-\Delta)$ ($\Delta = 10$, $\mu_1=5$ and $\mu_2=2.5$). Inset: The variance $\overline{Q^2}-\left(\overline{Q}\right)^2$ vs. time for classical oscillators with a Lorentzian distribution of frequencies. The shared initial state is $Q_0 = 25$ and $P_0=100$. Parameters are the same as in the main figure except for $\mu_1=2$ and $\mu_2=5$.

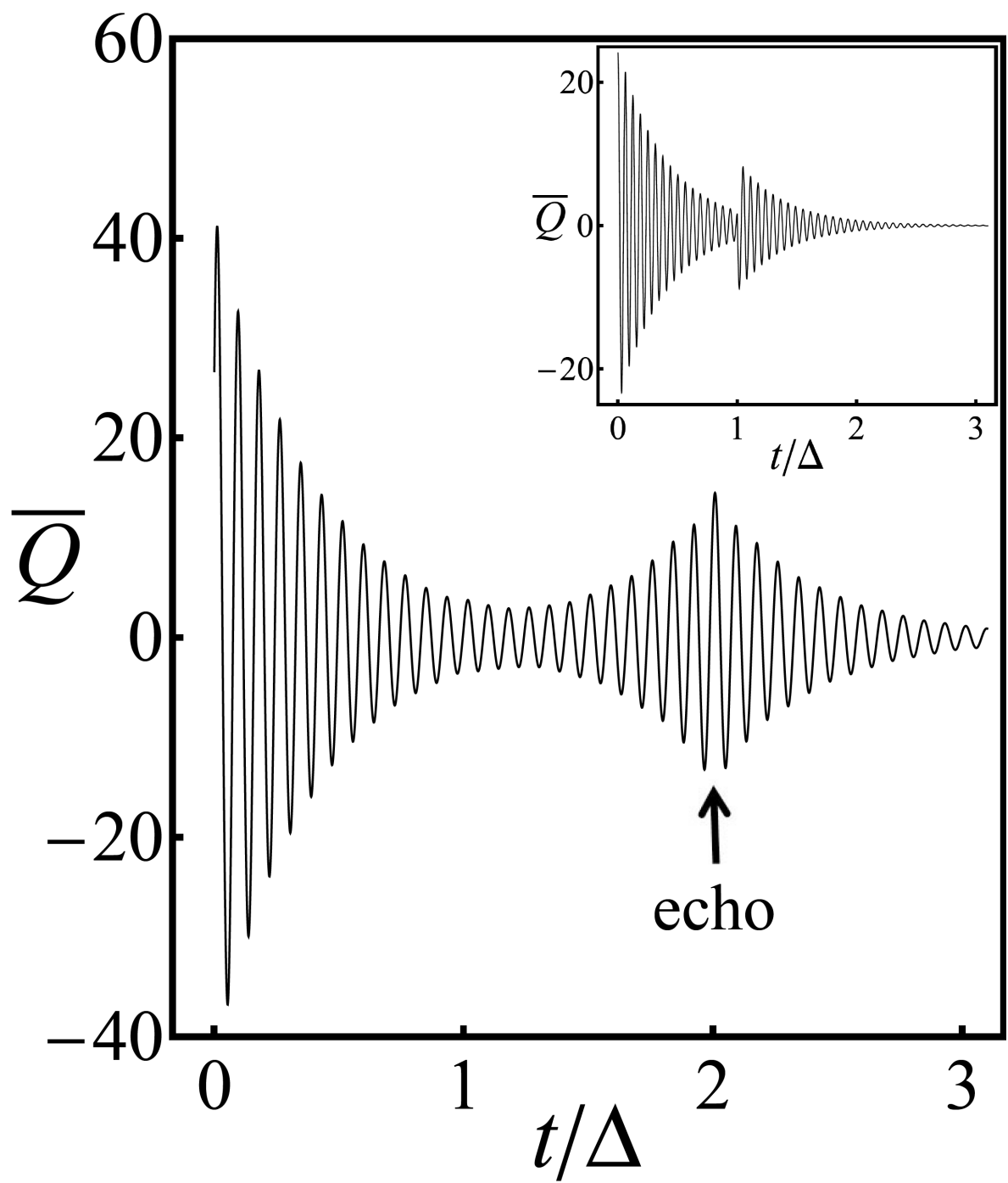

FIGURE 1

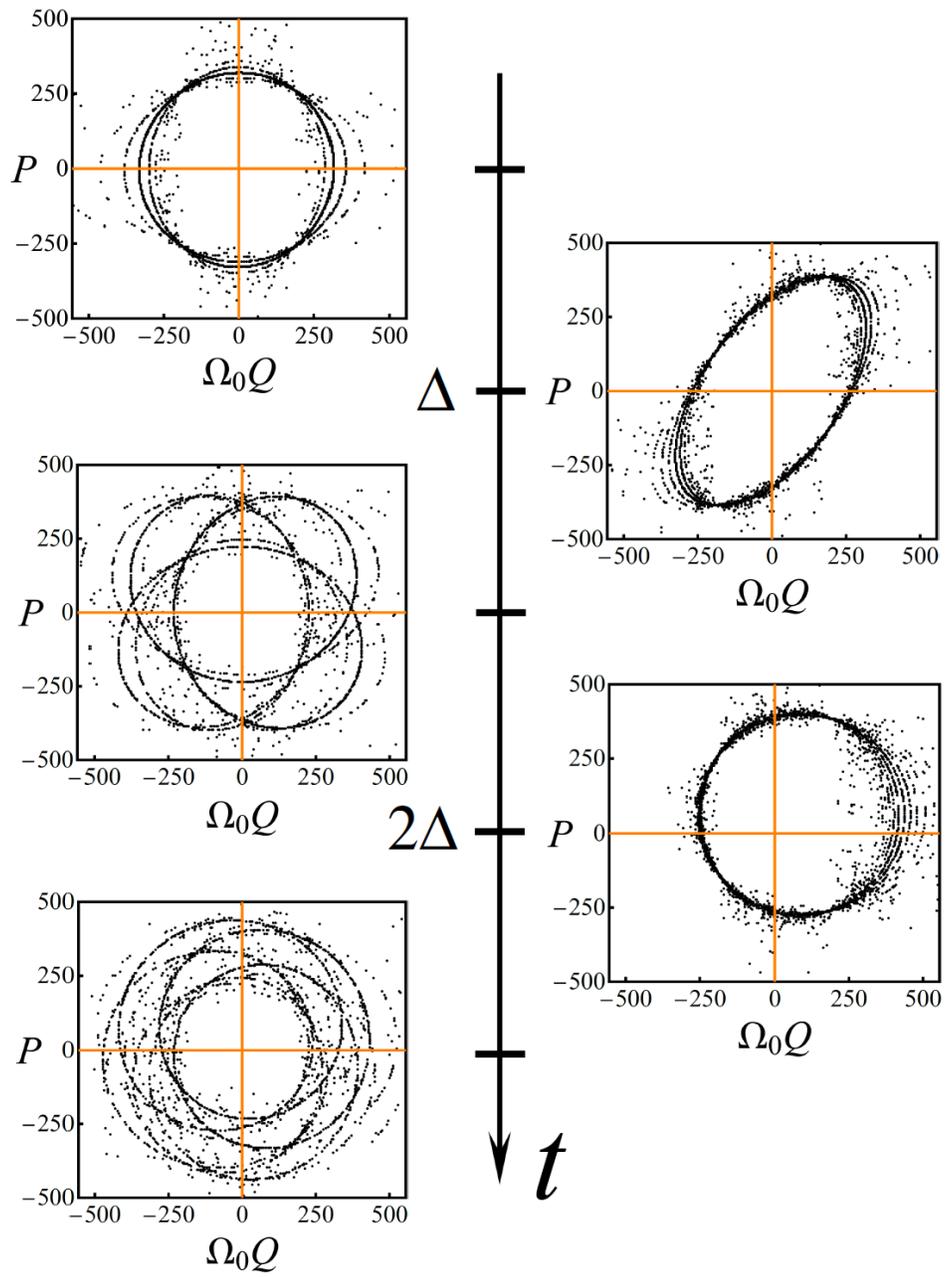

FIGURE 2

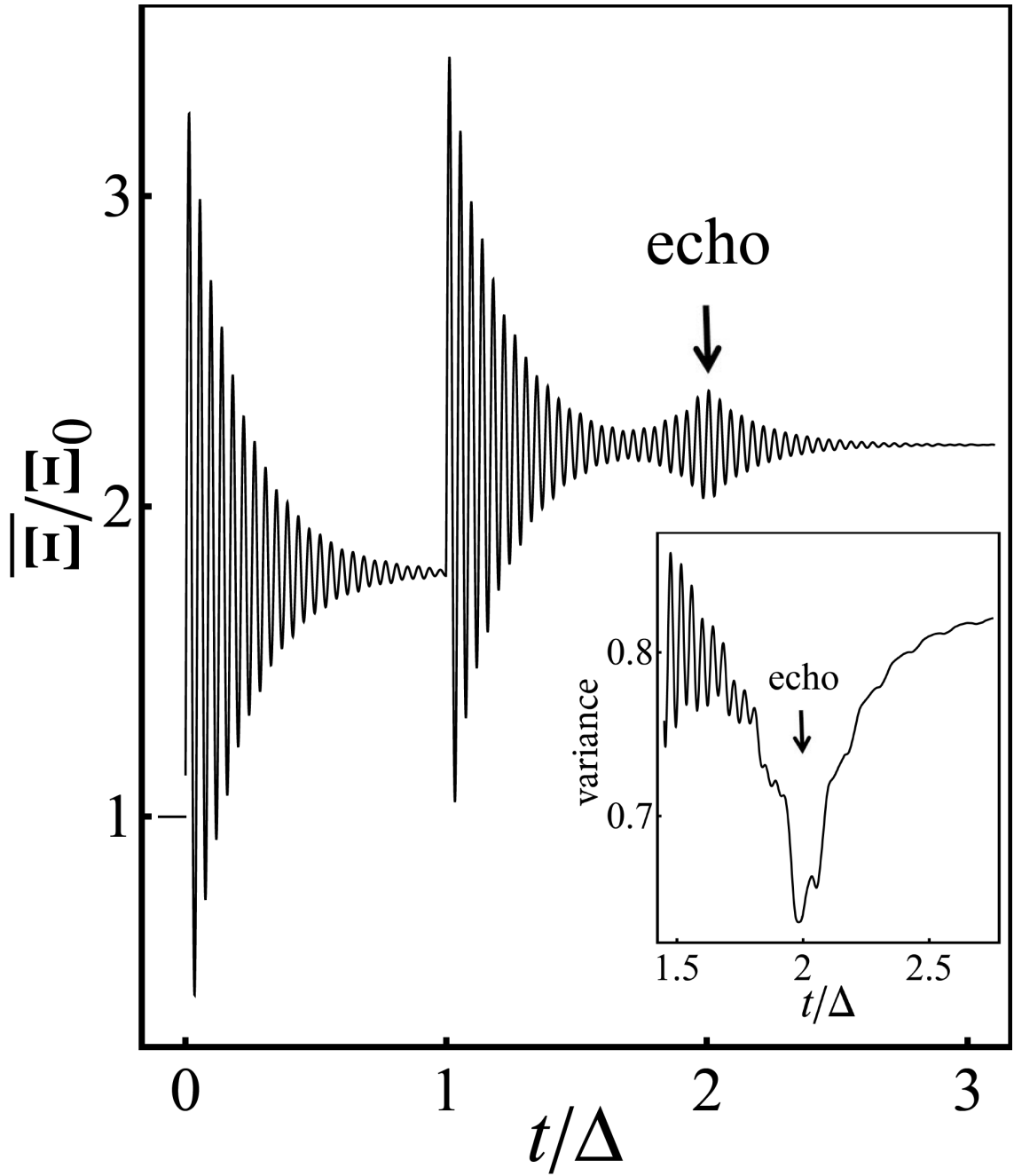

FIGURE 3